\documentclass[fleqn,10pt]{SelfArx}
\usepackage{lipsum}
\usepackage{natbib}

\definecolor{color1}{RGB}{0,0,90} 
\definecolor{color2}{RGB}{0,20,20} 

\usepackage{hyperref} 
\hypersetup{hidelinks,colorlinks,breaklinks=true,urlcolor=color2,citecolor=color1,linkcolor=color1,bookmarksopen=false,pdftitle={Title},pdfauthor={Author}}

 \def\gsim{\lower.4ex\hbox{$\;\buildrel >\over{\scriptstyle\sim}\;$}}
 \def\lsim{\lower.4ex\hbox{$\;\buildrel <\over{\scriptstyle\sim}\;$}}

\JournalInfo{Solar-Terrestrial Physics, 2017. Vol. 3. Iss. 1, pp. 19--25}
\Archive{}
\PaperTitle{CORRELATION PLOTS OF THE SIBERIAN RADIOHELIOGRAPH}

\Authors{S.V. Lesovoi\textsuperscript{1}*, V.S. Kobets\textsuperscript{1}}

\affiliation{\textsuperscript{1}\textit{Institute of Solar-Terrestrial Physics SB RAS, Irkutsk, Russia}}
\affiliation{*\textbf{E-mail}: svlesovoi@gmail.com}
\affiliation{**\textbf{www}: 
\href{http://badary.iszf.irk.ru/srhCorrPlot.php}{badary.iszf.irk.ru/srhCorrPlot.php}}

\newcommand{\keywordname}{Keywords}

\Abstract{The Siberian Solar Radio Telescope [\cite{Grechnev2003}] is now being upgraded. The upgrading is aimed at providing the aperture synthesis imaging in the 4–8 GHz frequency range [\cite{Lesovoi2012}, \cite{Lesovoi2014}] instead of the single-frequency direct imaging due to the Earth rotation. The first phase of the upgrading is a 48-antenna array — the  Siberian  Radioheliograph. One  type  of  radioheliograph  data  represents  correlation  plots\textsuperscript{**}. In evaluating the covariance of two-level signals, these plots are sums of complex correlations, obtained for different antenna pairs. Bearing in mind that correlation of signals from an antenna pair is related to a spatial frequency, we can say that each value of the plot is an integral over a spatial spectrum. Limits of the integration are defined by the task. Only high spatial frequencies are integrated to obtain dynamics of compact sources. The whole spectrum is integrated to reach maximum sensitivity. We show that the covariance of two-level variables up to Van Vleck correction is a correlation coefficient of these variables.}

\begin{document}
\flushbottom 

\maketitle 


\thispagestyle{empty} 

\section*{Introduction}
\begin{figure}
\centering{\includegraphics[width=\linewidth]{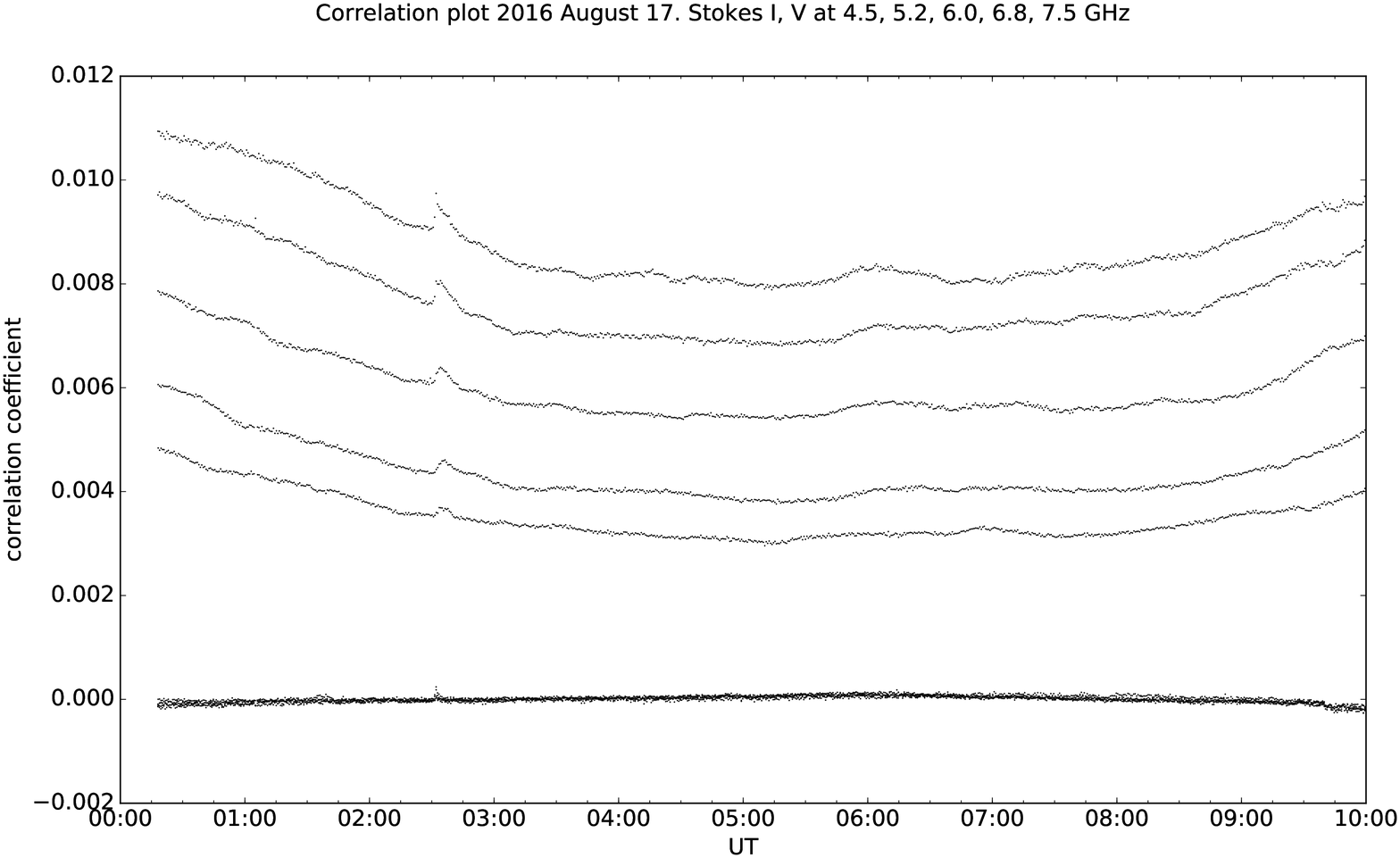}}
\caption{Example of daily regular data from the Siberian Radioheliograph. }
\label{corrPlot}
\end{figure}

The Siberian Solar Radio Telescope (SSRT) uses the Earth rotation and frequency scanning in a frequency band of 2$\%$ of the operating frequency for solar imaging [\cite{Grechnev2003}]. This approach limits temporal resolution to the time of passage of the Sun through the beam pattern. In addition, the SSRT receiving system consists of single frequency antennas and feeds. To solve current problems of solar-terrestrial physics, a sufficiently high temporal resolution and wide bandwidth are required. This would be possible only if, instead of the direct imaging, we use the Fourier synthesis. The first phase of the SSRT upgrading – a 48-antenna radioheliograph – has begun routine observations. [\cite{Lesovoi2012}, \cite{Lesovoi2014}]. The operating mode of the radioheliograph involves Fourier synthesis of solar images every 5 seconds at five frequencies in a range 4–8~GHz. Along with full-disk solar images, so-called correlation plots are of interest. A correlation plot of the Siberian Radioheliograph is shown in figure~\ref{corrPlot}. Data of this type are very convenient for solar activity monitoring and are characterized by very high sensitivity. Perhaps for the first time these data were introduced  into  practice of solar observations by the Nobeyama Radioheliograph team\footnote{ solar.nro.nao.ac.jp/norh/html/cor\_plot}. The purpose of this study is to explain in detail what the correlation plot is and what we can expect from data of this type. First we substantiate the statement that each point of correlation plot represents the sum of correlation coefficients, and then show the relationship between the correlation plots and the flux density, and explain their daily trends. 

\section*{CROSS-CORRELATION OF TWO-LEVEL SIGNALS}
Let us consider how a two-element correlation interferometer works. It is designed to measure the spatial coherence function of the field incident on an antenna pair. In other words, suppose the unknown is the time-dependent covariance of signals from an antenna pair. This covariance is measured by the correlator which is part of the interferometer. We represent signals from the antennas $k, l$ as the sum of a signal from a given object $s(t)$ and noises $n_k(t)$ and $n_l(t)$, generated by the receiving system $x_k(t) = \alpha(t) s(t) + n_k(t)$, $x_l(t) = \alpha(t) s(t) + n_l(t)$. The coefficient $\alpha$ shows the time variation of a signal from the given object. For example, when observing a solar flare, $\alpha$ is linearly related to the time variation of flux density. Suppose the signals $x_k, x_l$ - are real random variables with normal distribution, zero mean, and variances $\sigma_{x_k}^2, \sigma_{x_l}^2$; hen the joint density of probability of $x_k, x_l$ is defined as: 

\begin{equation}
\begin{split}
f(x_k, x_l)=\frac{1}{2\pi\sqrt{\left(1-\rho^2\right)}} e^{-\frac{1}{2\left(1-\rho^2\right)}\left[\frac{x_k^2}{\sigma^2_{x_k}} - \frac{2\rho x_k x_l}{\sigma_{x_k}\sigma_{x_l}} + \frac{x_l^2}{\sigma^2_{x_l}}\right]},
\end{split}
\end{equation}
where $\rho$ - is the correlation coefficient between $x_k$ and $x_l$.

Each of the $x_k, x_l$ variables contains $s$, therefore the covariance $\left<x_k x_l\right>$ differs from $0$and depends on $\alpha(t)$. As $\alpha(t)$ changes, the correlation coefficient varies as follows:
\begin{equation}
\rho(t) = \frac{\left<x_k x_l\right>}{\sigma_{x_k}\sigma_{x_l}}=\frac{\alpha^2(t)}{\alpha^2(t) + 1}.
\label{corrCoefAlpha}
\end{equation}
In the general case, the spatial coherence function is complex due to the asymmetry of the extended source. Accordingly, before evaluating the covariance, the analytic signals $z_k = x_k + i y_k,z_l = x_l + i y_l$, should be obtained from real signals $x_k, x_l$, where $y_k, y_l$ are related to $x_k, x_l$ by the Hilbert transform. The correlator estimates the $C_{kl}$ covariance, assuming that for analytic signals there holds 

$\left<x_k x_l\right> = \left<y_k y_l\right>$ and $\left<x_k y_l\right> = -\left<y_k x_l\right>$ ([\cite{BEN2016}]) and hence

\begin{equation}
\label{cov_def}
C_{kl} = \left<x_k x_l\right> + i\left<x_k y_l\right> = \left<z_k z^*_l\right>/2.
\end{equation}
The complex correlation coefficient of the random variables $z_k, z_l$  
\begin{equation}
\label{rho_def}
\rho_{kl} = \frac{\left<z_k z^*_l\right>}{\sigma_{z_k}\sigma_{z_l}}.
\end{equation}
Suppose $\hat{x}, \hat{y}$ are two-level signals derived from $x, y$. One-bit signal quantization is widely used in radioastronomy to reduce the correlator’s input data flow. At the output of the correlator, a signal is formed:
\begin{equation}
\label{corr_output}
N \hat{C}_{kl} =  \sum_{n=0}^{n=N-1}{\hat{x}_{kn}\hat{x}_{ln}} + i\sum_{n=0}^{n=N-1}{\hat{x}_{kn}\hat{y}_{ln}}.
\end{equation}
where $n$ is the number of the sample in the data collected over the accumulation time. Variance of the complex random variable $\sigma_{z}^2 = \left<x^2\right> + \left<y^2\right>$. In the analytical signal, variances of real and imaginary parts are equal, then $\sigma_{z}^2 = 2\left<x^2\right>$. In the two-level quantization of the random variable, its variance is $1$, then expression~\ref{rho_def} takes the form
\begin{equation}
\hat{\rho}_{kl} = \frac{\left<\hat{z}_k \hat{z}^*_l\right>}{2} = \hat{C}_{kl}.
\end{equation}

First consider the real part of expression~\ref{corr_output}. From Price’s theorem [\cite{Price1958}] it follows that
\begin{equation}
\frac{\partial\left<\hat{x}_k\hat{x}_l\right>}{\partial\rho_{xx}} = \left<\frac{\partial^2{\hat{x}_k \hat{x}_l}}{\partial x_k \partial x_l}\right>.
\end{equation}

The function that converts $x$ into $\hat{x}$ can be written as the difference of Heaviside functions: $\eta\left(x\right) - \eta\left(-x\right)$, the
derivative of which is $2\delta\left(x\right)$. Then, taking into account differentiation rules, the expression for derivative of covariance is written as
\begin{equation}
\frac{\partial\left<\hat{x}_k\hat{x}_l\right>}{\partial\rho_{xx}} = \left<\frac{\partial\hat{x}_k}{\partial x_k} \frac{\partial\hat{x}_l}{ \partial x_l}\right> = \left<4\delta\left(x_k\right)\delta\left(x_l\right)\right>.
\end{equation}
Bearing in mind the definition of the expectation for the function of random variable, we obtain:
\begin{equation}
\begin{split}
\frac{\partial\left<\hat{x}_k\hat{x}_l\right>}{\partial\rho_{xx}} = \int{\int{4\delta(z_k)\delta(x_l)f(x_k,x_l)\mathrm{d}x_k}\mathrm{d}x_l}=\\=\frac{2}{\pi\sqrt{\left(1-\rho_{xx}^2\right)}}.
\end{split}
\end{equation}
Whence it follows that:
\begin{equation}
\left<\hat{x}_k\hat{x}_l\right> = \frac{2}{\pi}\arcsin\left(\rho_{xx}\right).
\end{equation}
Inferring from $\left<x_k y_l\right>$, we get
\begin{equation}
\left<\hat{x}_k\hat{y}_l\right> = \frac{2}{\pi}\arcsin\left(\rho_{xy}\right).
\end{equation}
\begin{figure}
\centerline{\includegraphics[width=\linewidth]{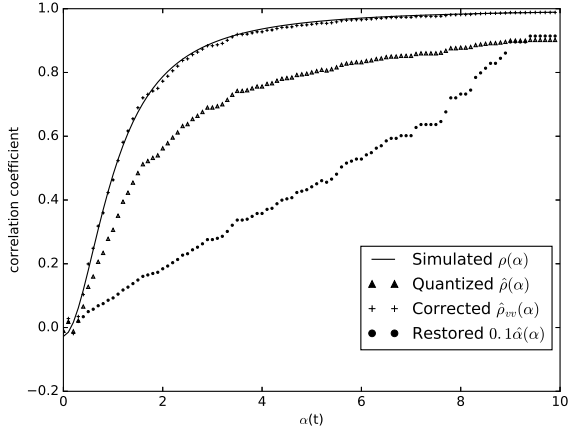}}
\caption{The response of the two-level correlator as a function of the signal/noise ratio for input signals: the solid curve is the result of simulation of the correlator response to nonquantized signals; triangles mark the result of correlator response simulation for two-level quantization; crosses indicate the Van Vleck correction; circles represent the $\alpha(t)$ dependence restored from the correlator response to two-level signals.}
\label{correctCorr}
\end{figure}
Thus, the estimated covariance of complex two-level variables, which is evaluated by the correlator up to the Van Vleck correction [\cite{VV1966}, \cite{BEN2016}], is equal to the correlation coefficient of original variables.
\begin{equation}
\rho_{kl} = \sin\left(\frac{\pi}{2}\left<\hat{x}_k\hat{x}_l\right>\right) + i\sin\left(\frac{\pi}{2}\left<\hat{x}_k\hat{y}_l\right>\right).
\end{equation}
The correlation plots in Figure~\ref{corrPlot} are averaged absolute values of correlation coefficients measured at certain frequencies:
\begin{equation}
\rho\left(t_n,\nu_m\right) = \frac{1}{KL}\sum_{k=0}^{k=K-1}{\sum_{l=0}^{l=L-1}{\left|\rho_{kl}\left(t_n,\nu_m\right)\right|}},
\end{equation}
where $K, L$ are numbers of antennas in the radioheliograph east-west and south arrays. Time dependence of quantized signal covariance~(\ref{corrCoefAlpha}) can yield a value linearly related to the time variation in flux density:
\begin{equation}
{\alpha\left(t_n, \nu_m\right)} = \sqrt\frac{\rho\left(t_n,\nu_m\right)}{1-\rho\left(t_n,\nu_m\right)}.
\end{equation}
The dependence restored in this way is shown in Figure~\ref{correctCorr}.

\section*{SENSITIVITY AND DIURNAL VARIATION OF THE CORRELATION PLOT}
To estimate the flux density corresponding to a given value of the correlation plot, we can use the fact that with small changes of the correlation coefficient (less than $0.6$) there is a linear relationship (Figure~\ref{correctCorr}) between them. Making use of microwave emission bursts with known flux density, we can show that  a $1\%$ change in the correlation corresponds to a flux density change by about $5-10$ sfu depending on the operating frequency. The time dependence of flux density can also be obtained from radioheliograph data. To do this, we should construct full-disk solar images for each moment of time and refer them to the brightness temperature scale. Once the images are presented in brightness temperatures, we can plot the density of a flux from a selected microwave source as a function of time by integrating brightness temperature over a chosen vicinity of the microwave source, as described in [\cite{KOCH2013}]. The expected sensitivity was estimated for the temperature of the receiving system at $2\cdot10^3 $~K and for the correlator efficiency at 0.8. Such efficiency of the two-level correlator is achieved through fivefold excess of sampling frequency over the Nyquist frequency [\cite{TMS2003}]. Temperature of the system is specified taking into account that the optical modulator, installed in each antenna, makes a noise contribution comparable to the level of signal from the Sun. The estimated radioheliograph sensitivity to flux density at an operating bandwidth of 10 MHz and an  accumulation time of 0.3~s is $5\cdot10^{-3}$~sfu, providing that the effective area of a single antenna is one half of the geometrical area. The sensitivity measured from fluctuations of the correlation plot is at least $10^{-2}$~sfu. 

\begingroup
\centerline{\includegraphics[width=\linewidth]{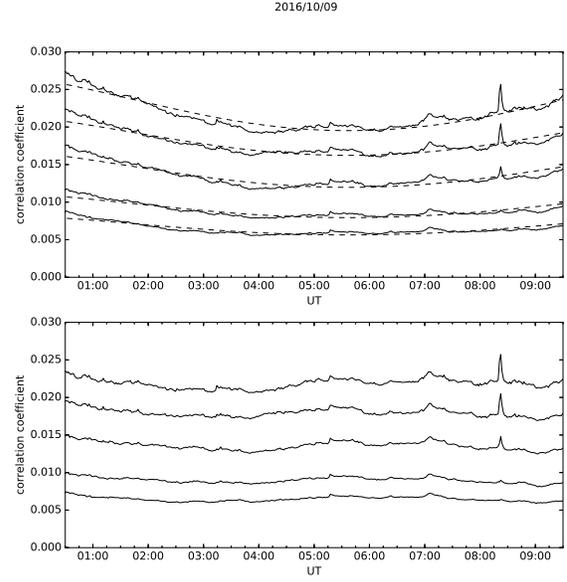}}
\captionof{figure}{Raw correlation plots of the Radioheliograph (top) and the same plots with compensation of daily trend (bottom).}
\label{corrPlotDailyCompensation}
\endgroup

The characteristic variation of the correlation plot during the day can be clearly seen in Figure~\ref{corrPlot}. The correlation coefficient decreases with hour angle. This follows from the fact that the degree of spatial coherence of the finite size source increases with decreasing distance between points where the coherence is measured. With increasing hour angle, the whole interferometer baseline effectively decreases due to reduction in projection of baselines of west-east antenna array. In other words, the contribution of smaller baselines increases with hour angle. The Van Cittert–Zernike theorem holds that the spatial coherence function measured for an antenna pair is proportional to the corresponding component of the spatial source spectrum [\cite{TMS2003}]. Given that the spatial spectrum of the finite size source decreases with increasing spatial frequency, the average correlation coefficient should increase with decreasing baseline. To approximate the diurnal variation of the correlation plot, we can use the dependence that is inverse to the average length of projections of radioheliograph baselines. The baseline projection for the antenna pair $m, n$ is calculated as follows [\cite{TMS2003}]
\begin{equation}
\begin{split}
b\left(m,n,t\right) = \\
=\left(
\begin{matrix}
\sin(h) & \cos(h) & 0\\
-\sin(\delta)\cos(h) & \sin(\delta)\sin(h) & \cos(\delta)\\
\cos(\delta)\cos(h) & -\cos(\delta)\sin(h) & \sin(\delta)\\
\end{matrix}
\right)\times\\
\times\left(
\begin{matrix}
r_x\left(m,n\right)\\
r_y\left(m,n\right)\\
r_z\left(m,n\right)\\
\end{matrix}
\right)
\end{split}
\end{equation}

where 
\begin{equation}
\begin{split}
\left(\begin{matrix}r_x\left(m,n\right)\\r_y\left(m,n\right)\\r_z\left(m,n\right)\end{matrix}\right) =
\left(
\begin{matrix}
-\sin(\phi) & 0 & \cos(\phi)\\
0 & 1 & 0\\
\cos(\phi) & 0 & \sin(\phi)\\
\end{matrix}
\right)\times\\
\times\left(
\begin{matrix}
(192.5 - m)d\\
(n - 64.5)d\\
0\\
\end{matrix}
\right)
\end{split}
\end{equation}
Here $h, \delta$ is the hour angle and declination, $d$ is the minimum distance between antennas, $m, n$ is the SSRT’s numbering of antennas: $192..177$, $49..80$ for south and west-east directions respectively. The value 
\begin{equation}B(t) =\\ \sum_{m=192}^{m=177}\sum_{n=49}^{n=80}\sqrt{b(m,n,t)_x^2 + b(m,n,t)_y^2}\end{equation}
can be used to approximate the diurnal variation of the correlation plot. Figure~\ref{corrPlotDailyCompensation} exemplifies the approximation of the diurnal variation of the correlation plot by $1/B(t)$. The correlation plot asymmetry is caused by the fact that at negative hour angles (in the morning) the antennas capture the background emission from forest. The effect of background emission is adequately compensated by a small positive shift of the local noon when calculating the diurnal variation of the correlation plot.

\section*{CONCLUSION}
The Siberian Radioheliograph’s correlation plots representing the sums of correlation coefficients or the integral over the spatial spectrum of solar image are very informative in the context of the investigation into the dynamics of microwave emission from the Sun. Although these plots are not linearly related to the time variation of flux density, for compact sources and not very large correlation coefficients (less than $0.6$, see Figure~\ref{correctCorr}) we can consider that a $1\%$ increment of correlation corresponds to an increment of about $5-10$~sfu for the operating frequency in the range $4-8$~GHz. The  sensitivity of the correlation plots in flux density is as high as $10^{-2}$~sfu for compact sources. This allows us to explore previously inaccessible weak bursts of microwave emission, both individual and coming before solar flares.

This work was supported by FASO under the project “Investigation into extremely weak solar activity in the microwave range” performed at SSRT, Fundamental Research Program 7 of the RAS Presidium “Experimental and theoretical studies of objects in the solar system and planetary systems of stars. Transitional and explosive processes in astrophysics”, and RFBR grant No. 15-02-01089 A.

\end{document}